# A Holistic Framework for Robust Bangla ASR and Speaker Diarization with Optimized VAD and CTC Alignment

Zarif Mahir
*Computer Science and Engineering*
*Bangladesh University of Engineering and Technology*
Dhaka, Bangladesh
zarifmahir21@gmail.com

Zarif Ishmam
*Computer Science and Engineering*
*Bangladesh University of Engineering and Technology*
Dhaka, Bangladesh
zarif.ishmam9@gmail.com

Shafnan Wasif
*Computer Science and Engineering*
*Bangladesh University of Engineering and Technology*
Dhaka, Bangladesh
shafnanwasif@gmail.com

Md. Ishtiak Moin
*Computer Science and Engineering*
*Bangladesh University of Engineering and Technology*
Dhaka, Bangladesh
zarifishtiak@gmail.com

*Abstract*—Despite being one of the most widely spoken languages globally, Bangla remains a low-resource language in the field of Natural Language Processing (NLP). Mainstream Automatic Speech Recognition (ASR) and Speaker Diarization systems for Bangla struggles when processing long-form audio exceeding 30-60 seconds. This paper presents a robust framework specifically engineered for extended Bangla content by leveraging preexisting models enhanced with novel optimization pipelines for the DL Sprint 4.0 contest. Our approach utilizes Voice Activity Detection (VAD) optimization and Connectionist Temporal Classification (CTC) segmentation via forced word-alignment to maintain temporal accuracy and transcription integrity over long durations. Additionally, we employed several fine-tuning techniques and preprocessed the data using augmentation techniques and noise removal. By bridging the performance gap in complex, multi-speaker environments, this work provides a scalable solution for real-world, long-form Bangla speech applications.

## I. Introduction

Bangla is the seventh most spoken native language in the world, with over 270 million speakers across the Indian subcontinent and a vast global diaspora[1]. Despite its linguistic prominence, Bangla remains categorized as a low-resource language in the context of Natural Language Processing (NLP) and Automatic Speech Recognition (ASR)[2]. While recent advancements in Deep Learning and Transformer-based architectures have improved short-snippet transcription, a critical performance bottleneck persists: long-form audio processing.

Most state-of-the-art ASR models for Bangla are trained and evaluated on segmented, high-quality clips typically under 30 seconds[3]. However, real-world applications—such as podcast transcription, judicial proceedings, and broadcast media—require systems capable of handling continuous audio streams spanning several minutes or hours. In these extended contexts, standard models often suffer from "hallucinations", catastrophic forgetting, or loss of temporal alignment, leading to a significant spike in Word Error Rate (WER)[4].

The challenge is twofold. First, long-form ASR requires precise segmentation to prevent the encoder from being overwhelmed by sequence length. Second, Speaker Diarization (SD)—the task of "who spoke when"—becomes increasingly complex as the number of speakers and the duration of the audio increases. Existing Bangla datasets rarely provide the necessary metadata for robust diarization in long-form, multi-speaker environments.

In this paper, utilizing the provided dataset for DL Sprint 4.0[5], we address these limitations by proposing a specialized pipeline engineered specifically for long-form Bangla content. Our framework enhances both transcription and speaker diarization through five core contributions:

1) **Audio Preprocessing and Noise Removal**: Employing advanced source separation techniques [6]
2) **Dynamic Data Augmentation**: Applying variable gain adjustments and dynamic audio augmentation during training to improve model robustness across diverse, real-world recording environments.
3) **VAD Optimization**: Implementing a refined Voice Activity Detection (VAD) layer to accurately identify and filter non-speech segments prior to transcription.
4) **CTC Alignment and Segmentation**: Utilizing Connectionist Temporal Classification (CTC) via forced word-alignment[7] to maintain precise temporal boundaries, allowing for intelligent audio chunking that preserves continuous speech context.
5) **Integrated Diarization**: Developing a unified framework that combines fine-tuned segmentation and embedding-based clustering to execute precise multi-speaker labeling in tandem with ASR.

## II. LITERATURE REVIEW

The development of robust Automatic Speech Recognition (ASR) and Speaker Diarization (SD) systems for Bengali has evolved from traditional statistical methods to deep learning architectures. However, long-form, multi-speaker audio processing remains challenging in low-resource settings.

### A. Advancements in Bengali Automatic Speech Recognition

Early Bengali ASR relied on Hidden Markov Models (HMM) and Gaussian Mixture Models (GMM) using read speech or broadcast news corpora[8]. These models struggled with Bengali's morphological complexity and dialectal variance. Crowdsourced datasets like Bengali Common Voice enabled deep learning shifts, such as Wav2Vec 2.0[9]. Recently, Transformer-based models like OpenAI's Whisper have shown strong zero-shot and fine-tuning performance. Fine-tuned Whisper models support tasks like clinical translation and speech-to-sign language generation[10]. However, Whisper is limited to 30-second inputs, leading to hallucinations, forgetting, and drift in long-form audio.

### B. Long-Form Processing and Temporal Alignment

Extended audio requires advanced segmentation and alignment. Traditional sliding-window methods often cut words, disrupting context for Transformer decoders. Recent work integrates Connectionist Temporal Classification (CTC) for forced word-level alignment, enabling precise timestamp extraction and boundary-respecting chunking to preserve linguistic integrity.

### C. Speaker Diarization in Low-Resource Contexts

Speaker Diarization involves Voice Activity Detection (VAD), segmentation, embedding extraction, and clustering. Toolkits like `pyannote.audio` standardize these for high-resource languages[11]. Bengali diarization resources are scarce, but initiatives like Bengali-Loop and DL Sprint 4.0 provide annotated long-form datasets[5]. Out-of-the-box models show high Diarization Error Rates (DER) due to noise, overlaps, and environments. Research now focuses on hybrid approaches with vocal source separation and fine-tuning for scalable pipelines.

## III. METHODOLOGY

### A. Long-form ASR

Our proposed approach is structured into two primary phases: robust dataset preprocessing via CTC-based forced alignment, and acoustic model fine-tuning utilizing a pretrained Whisper architecture. The complete pipeline is implemented using Python, leveraging the Hugging Face `transformers` and `datasets` libraries, alongside `ctc-forced-aligner` for precise temporal bounding.

a) *Dataset Preprocessing and Forced Alignment:* To address the challenges of long-form Bangla speech, we utilize datasets such as the DL Sprint 4.0 (Bengali Long-form Speech Recognition) corpus. Raw audio sequences must be precisely segmented to accommodate the fixed input constraints of transformer-based ASR models without truncating contextual speech.

**Data Normalization:** Audio files (`.wav`) are paired with their corresponding transcripts (`.txt`). To ensure textual consistency, all transcripts undergo Unicode Normalization Form Canonical Composition (NFC) and rigorous whitespace stripping to remove orthographic anomalies.

**CTC Forced Alignment:** We utilize the multilingual MMS-300M forced aligner, specifically configured to its Bengali language head. The model is loaded in float16 precision on GPU hardware to accelerate emission generation. For each audio-transcript pair, the process is as follows:

1) Emission Generation: The alignment model's encoder generates frame-level CTC emissions.
2) Textual Preprocessing: Transcripts are romanized to map the Bengali script to the acoustic phonetic representations expected by the aligner.
3) Word-Level Alignment: The system computes Viterbi alignments, extracting precise start and end timestamps for each localized word.
4) Error Handling: A fallback mechanism is implemented to filter or flag instances of mismatched or empty alignments caused by extreme background noise or untranscribed speech.

**Chunking and Segmentation:** Utilizing the extracted word-level timestamps, the continuous audio stream is intelligently segmented into discrete chunks of strictly under 30 seconds. This step ensures that word boundaries are preserved —preventing cuts in the middle of a spoken word—while

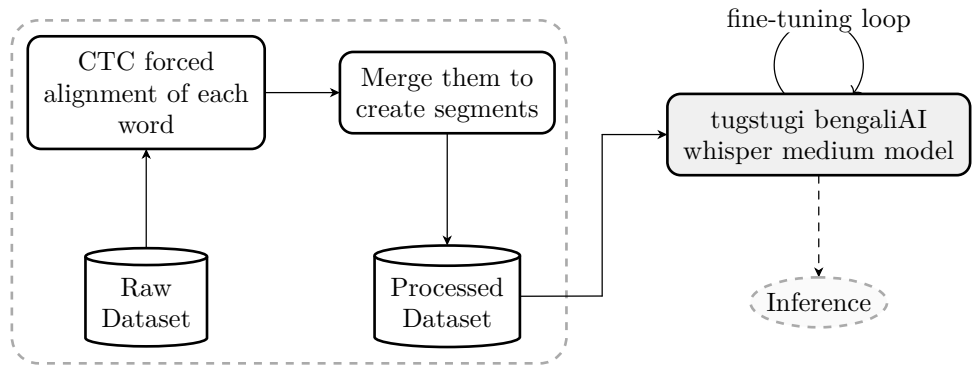

Fig. 1. Long-form ASR fine-tuning pipeline

strictly adhering to the maximum input length of the Whisper architecture.

b) *Acoustic Model Fine-Tuning:* For the transcription engine, we fine-tune the `bengaliAI/ tugstugi_bengaliai-asr_whisper-medium` [12] model using our newly aligned and segmented dataset. All training procedures were executed on an NVIDIA A100 GPU.

**Data Preparation:** Prior to feature extraction, all audio chunks are strictly resampled to 16 kHz to match the Whisper encoder's expected temporal resolution. Transcripts are tokenized using the native Whisper processor to map Bengali text to the corresponding byte-level BPE tokens.[13]

TABLE I
Summary Statistics of the Bengali ASR Chunked Dataset

| Attribute | Description |
| --- | --- |
| Dataset Name | `zarifmahir21/ bengali-asr-chunked` |
| Primary Task | Automatic Speech Recognition (ASR) |
| Total Utterances | 13,547 |
| Total Duration | ≈ 158 Hours |
| Audio Format | WAV (Mono) |
| Sampling Rate | 16 kHz |
| Chunking Strategy | Segmented to < 30s (Whisper limit) |

**Training Protocol:** The model is trained focusing on Bengali token distributions using a batch size of 16. To manage distributed training and optimize GPU memory utilization, we employ the Hugging Face `Accelerate` library. Model performance is continuously evaluated across epochs by calculating the Word Error Rate (WER) using the `jiwer` metric, as shown in Table II.

TABLE II
Training and Validation Metrics for Whisper Fine-Tuning

| Step | Training Loss | Validation Loss | WER |
| --- | --- | --- | --- |
| 190 | 0.939129 | 0.195895 | 26.391037 |
| 380 | 0.551614 | 0.165256 | 22.931987 |
| 570 | 0.244060 | 0.167897 | 22.711062 |

*B. Bangla Long-form Speaker Diarization*

To address the highly complex acoustic environments of real-world Bengali audio—which frequently feature background noise, overlapping speech, and varying recording volumes—we engineered a sophisticated three-stage Curriculum Learning pipeline. Rather than training a model natively on a static dataset, our approach progressively guides the model from broad acoustic adaptation to highly specific vocal refinement, culminating in a robustness stress-test using dynamic data augmentation.

a) *Stage 1: Base Adaptation (Real-World Acoustics):* The initial phase focuses on adapting a multilingual baseline (`pyannote/segmentation-3.0`) to the specific phonetic cadence and overlapping conversational patterns of the Bengali language.

**Data Preparation:** Raw audio files (`.wav`) from the training corpus are aligned with their respective CSV annotations, converted into PyAnnote's native Rich Transcription Time Mark (RTTM) format.

**Noisy Fine-Tuning:** The model is trained on these unaltered, raw audio chunks. In this phase, the network learns to identify fundamental Bengali speech characteristics while simultane-ously navigating the natural background noise, static, and music inherent to the original recordings. The best-performing checkpoint is versioned and exported for subsequent refinement.

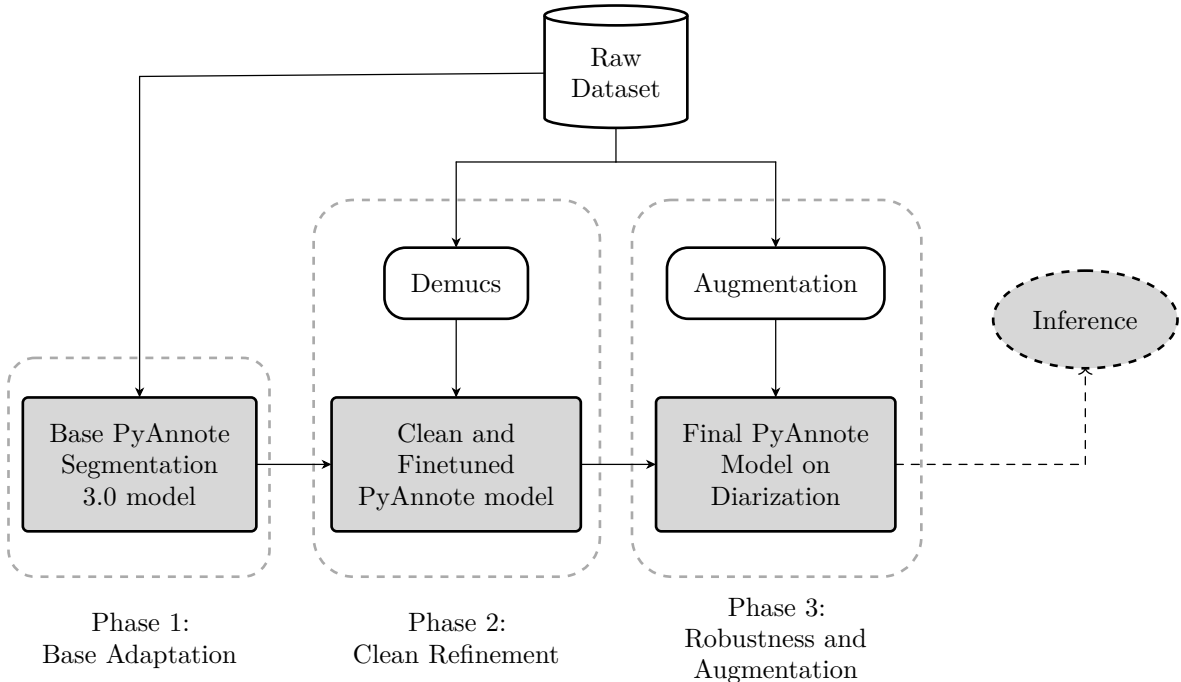

Fig. 2. Three-phase Speaker Diarization fine-tuning pipeline

b) *Stage 2: Clean Refinement (Source Separation):* While Phase 1 builds a generalized understanding of the acoustic environment, Phase 2 isolates the distinct mathematical signatures—or speaker embeddings—of individual voices.

**Vocal Isolation:** We systematically process the entire training directory using the Demucs hybrid Transformer-based architecture (`htdemucs --two-stems=vocals`). This effectively strips away background noise and non-speech artifacts, outputting pure vocal tracks.

**Audio Standardization:** The isolated vocal tracks are converted to a single monaural channel and strictly resampled to 16 kHz to align with the PyAnnote ecosystem's input requirements.

**Clean Fine-Tuning:** Utilizing the specialized model weights derived from Stage 1, we resume training exclusively on the Demucs-cleaned audio. By removing background distractions, the network focuses entirely on perfecting boundary detection and speaker differentiation.

c) *Stage 3: Robustness and Dynamic Augmentation:* Models trained exclusively on source-separated audio risk overfitting to the artificial cleanliness and specific volume levels of the Demucs output. To mitigate this brittleness, the final phase introduces controlled stochastic distortions.

**Dynamic Dataloader Augmentation:** We integrate `torch_audiomentations` to apply on-the-fly volume transformations. Specifically, a Gain augmentation (ranging from $-6.0$ dB to $+6.0$ dB) is applied with a 40% probability ($p = 0.4$) to any given 5.0-second chunk during the training loop.

**Custom Wrapper Implementation:** Due to the strict data formatting constraints of PyAnnote's internal architecture, we implemented a custom `ManualExternal` PyTorch module. This wrapper safely catches the augmented audio tensors and seamlessly bundles them with their corresponding target labels, ensuring stability during the forward pass.

*Training Protocol and Hyperparameters:* All fine-tuning stages were executed using PyTorch Lightning on GPU hardware (particularly the Tesla T4 GPUs on Kaggle). The `SpeakerDiarization` task was configured to process fixed audio chunks, optimizing for cross-entropy loss. The complete hyperparameter configuration utilized across the advanced training phases is detailed in Table III.

TABLE III
CURRICULUM LEARNING HYPERPARAMETERS FOR SPEAKER DIARIZATION

| Parameter | Phase 1 | Phase 2 | Phase 3 (Augmentation) |
|---|---|---|---|
| Chunk Duration | 5.0 s | 5.0 s | 10.0 s |
| Batch Size | 32 | 32 | 32 |
| Learning Rate | $5 \times 10^{-5}$ | $5 \times 10^{-5}$ | $5 \times 10^{-5}$ |
| Epochs | Variable | Variable | 20 |
| Audio Source | Raw Audio | Demucs Vocals | Demucs Vocals |
| Augmentation Protocol | None | None | Dynamic Gain ($\pm 6.0$ dB, $p = 0.4$) |
| Loss Function | Cross-Entropy | Cross-Entropy | Cross-Entropy |

## IV. CONCLUSION

Modern Transformer-based ASR models typically utilize restricted context windows—often between 30 and 60, as computational complexity scales quadratically with sequence length. Consequently, specialized strategies are required to maintain accuracy in long-form ASR and speaker diarization. The following summary outlines our findings across these various approaches:

TABLE IV
ASR PERFORMANCE BENCHMARK ON BENGALI PUBLIC AND PRIVATE TEST SETS

| Model Architecture | Configuration | Public WER ↓ | Private WER ↓ | Time |
|---|---|---|---|---|
| Tugstugi | Fine-tuned | **0.21988** | **0.23585** | 4h |
| Tugstugi | Zero-shot | 0.36142 | 0.37871 | 4h |
| Bangla-ASR | Fine-tuned | 0.50047 | 0.54329 | 2.5-3h |
| Mozilla Large | Base | 0.63171 | 0.69726 | 12h |
| Whisper Large Turbo v3 | Zero-shot | 0.86594 | 0.88630 | 4-5h |

TABLE V
SPEAKER DIARIZATION PERFORMANCE (DER) ACROSS DIFFERENT TRAINING STRATEGIES

| Training Strategy | Public DER ↓ | Private DER ↓ |
|---|---|---|
| Normal Fine-tuning (Base) | 0.23147 | 0.31129 |
| Fine-tuning + Demucs Refinement | 0.21621 | 0.33454 |
| Fine-tuning + Data Augmentation | **0.21460** | **0.32663** |
| External Dataset + Augmentation | 0.22523 | 0.33982 |
| External Dataset (No Augmentation)[14] | 0.23163 | 0.33543 |

We optimized inference performance on Kaggle by utilizing dual T4 GPUs in parallel, reducing the total inference time for the fine-tuned `bengaliAI/tugstugi_bengaliai-asr_whisper-medium` model to approximately two hours.